\documentclass[superscriptaddres,notitlepage]{revtex4-1}

\usepackage[margin=1.0in]{geometry}
\usepackage{verbatim}
\usepackage{graphicx}
\usepackage{amsmath}
\usepackage{color}
\usepackage{setspace}

\begin{document}

\title{High-power femtosecond pulses without a modelocked laser}

\author{Walter Fu,$^{1,*}$
Logan G. Wright,$^{1}$
Frank W. Wise$^{1}$
}
\affiliation{
	$^{1}$School of Applied and Engineering Physics, Cornell University, Ithaca, New York 14853, USA
}
\affiliation{
	$^{*}$Corresponding author: wpf32@cornell.edu
}

\begin{abstract}
We demonstrate a fiber system which amplifies and compresses pulses from a gain-switched diode.  A Mamyshev regenerator shortens the pulses and improves their coherence, enabling subsequent amplification by parabolic pre-shaping.  As a result, we are able to control nonlinear effects and generate nearly transform-limited, 140-fs pulses with 13-MW peak power---an order-of-magnitude improvement over previous gain-switched diode sources.  Seeding with a gain-switched diode results in random fluctuations of 2\% in the pulse energy, which future work using known techniques may ameliorate.  Further development may allow such systems to compete directly with sources based on modelocked oscillators in some applications while enjoying unparalleled robustness and repetition rate control.
\end{abstract}

\maketitle

Fiber lasers are becoming increasingly widespread in scientific and industrial settings.  In contrast with their solid-state counterparts, fiber systems benefit from a compact, robust format and diffraction-limited beam quality even at high average powers.  However, the majority of their impact has to-date been in long-pulse or continuous-wave applications.  While few fiber systems can generate millijoule-scale, ultrafast pulses \cite{Galvanauskas2001}, microjoule-level pulses obtainable by amplifying a modelocked fiber oscillator hold advantages for imaging, precision machining, and ultrafast measurement.  Nevertheless, such systems have yet to find widespread adoption outside of research labs.  Their limited impact stems from several remaining practical obstacles.  Maintaining stable modelocking in the face of environmental perturbations is an ongoing engineering problem.  Separately from this issue, modelocked oscillators are constrained to operate at their fundamental repetition rate.  This loss of flexibility can be limiting for applications where pulses must be synchronized with scanning optics or other components, or where both average and peak power need to be optimized in tandem.  Key examples include nonlinear microscopy and micromachining.  In both cases, tailoring the temporal pulse pattern permits not only greater throughput, but also entirely new imaging and material processing modalities \cite{Farrar2011,Kerse2016}.

Gain-switched diodes (GSDs) offer a way to overcome these challenges.  As integrated optoelectronic devices, they are robust against environmental fluctuations.  They also can be electronically triggered to produce arbitrary repetition rates or more complicated pulse trains.  From an applications standpoint, this is a key advantage over modelocked oscillators.  However, GSDs also have a number of disadvantages.  Ultrafast (fs) processes in passively modelocked laser cavities produce highly coherent pulses with fs-scale durations.  By contrast, GSDs form pulses from noise using far slower processes.  The pulses therefore exhibit much greater interpulse amplitude and phase noise, as well as higher intrapulse phase noise; consequently, pulses are much longer (10-100 ps) and cannot be directly compressed to the transform limit.  As a result, despite offering several unique capabilities, GSDs have thus far had almost no impact in fields demanding high peak powers and femtosecond-scale pulses.

Much research has focused on making GSDs more attractive as ultrafast sources.  Some of the earliest work in this area used soliton compression to obtain pulse durations of hundreds of femtoseconds \cite{Nakazawa1990} or even shorter \cite{Matsui1999}, often in conjunction with amplification or nonlinear pulse cleaning \cite{Ahmed1995}.  However, soliton effects ultimately limited these pulses to picojoule-scale energies.  Bypassing these constraints through chirped-pulse amplification enabled one early, standout result to achieve 2-nJ, 230-fs pulses, but required a specially engineered diode structure \cite{Galvanauskas1993}.  In the past few years, amplification of GSDs to microjoule energies has been demonstrated, but with picosecond-scale durations \cite{Chen2010,Heidt2013}.  Only recently have GSDs been amplified to such high energies at sub-picosecond durations \cite{Fang2016}.  A key component in this system was a nonlinear spectral-temporal filtering process \cite{Liu2016} similar to the well-known Mamyshev regenerator \cite{Mamyshev1998}.  While this stage was primarily intended to suppress amplified spontaneous emission (ASE), it had the side effect of shortening the pulses by a modest factor.  Subsequent amplification in the normally-dispersive regime permitted pulse compression to 0.6 ps with peak powers exceeding 1 MW.  Although this represented a new record for GSDs, the pulses remain too long for many ultrafast applications, and their large deviation from the transform-limited duration (>4x) indicates a lack of control over the pulse.  In addition, the system relies on a specially engineered diode, which emits high-quality seed pulses.

Here, we demonstrate a scheme for compressing and amplifying pulses from a GSD.  Starting with partially coherent, $\approx$10-ps pulses from a commercially-available GSD, we generate 140-fs pulses with energies of 2.4 $\mu$J.  These represent the shortest pulses from a GSD at comparable energies, and the highest peak power by an order of magnitude.  We achieve these results using a multi-stage system.  We first use a Mamyshev regenerator to shape the pulses from the GSD, at once compressing them and improving their coherence.  The clean pulses that we thereby obtain are then amenable to parabolic pre-shaping, followed by amplification and compression to femtosecond durations \cite{Pierrot2013}.  By using propagation regimes that effectively leverage nonlinear effects, we are able to compress the amplified pulses to near the transform limit.

\begin{figure}[htb]
\centering
\includegraphics[width=0.5\linewidth]{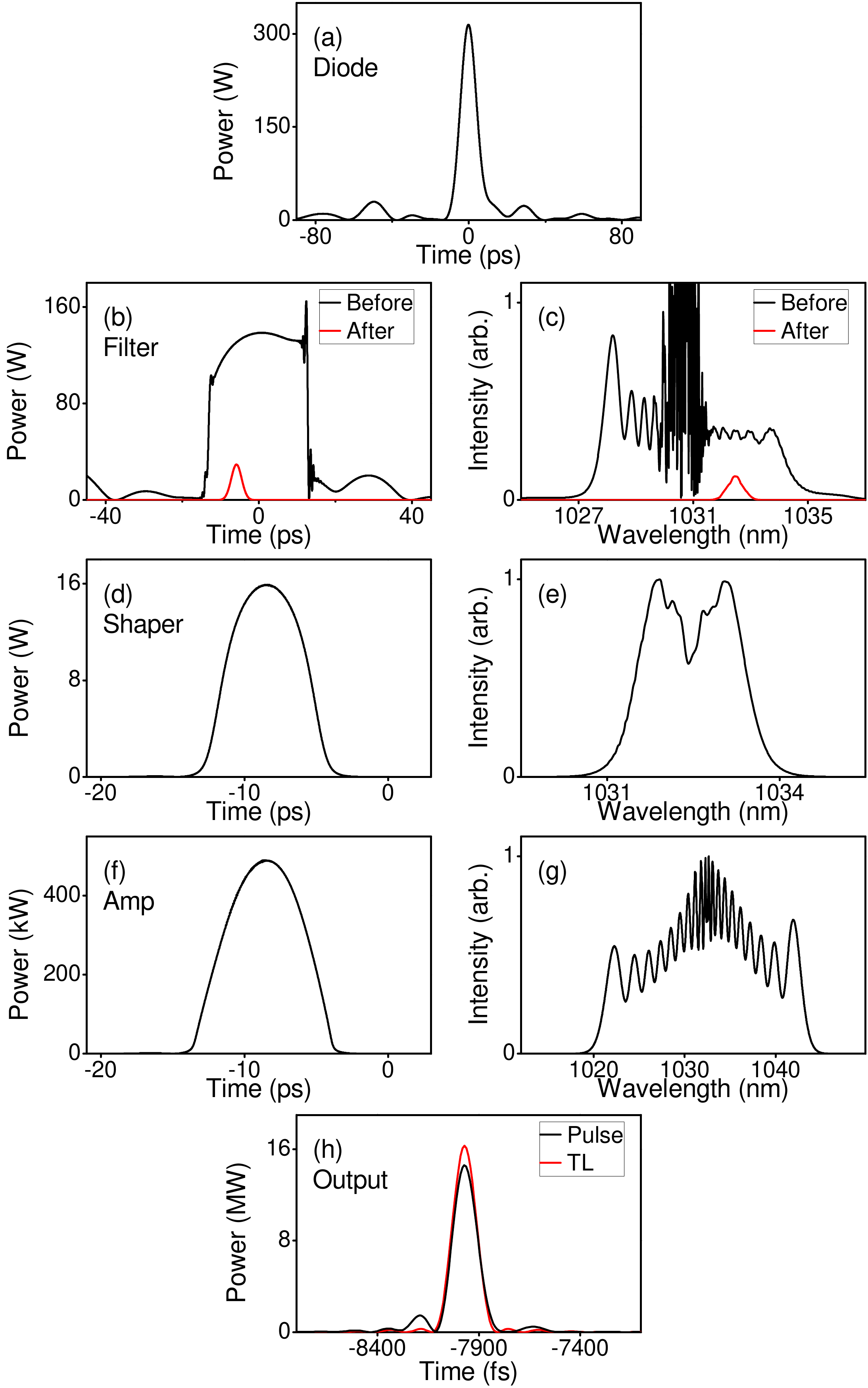}
\caption{Simulation results (a) at the output of the GSD, (b-c) before and after the filter, (d-e) after parabolic shaping, (f-g) after amplification, and (h) after dechirping.}
\label{fig:sim}
\end{figure}

\begin{figure}[!htb]
\centering
\includegraphics[width=0.6\linewidth]{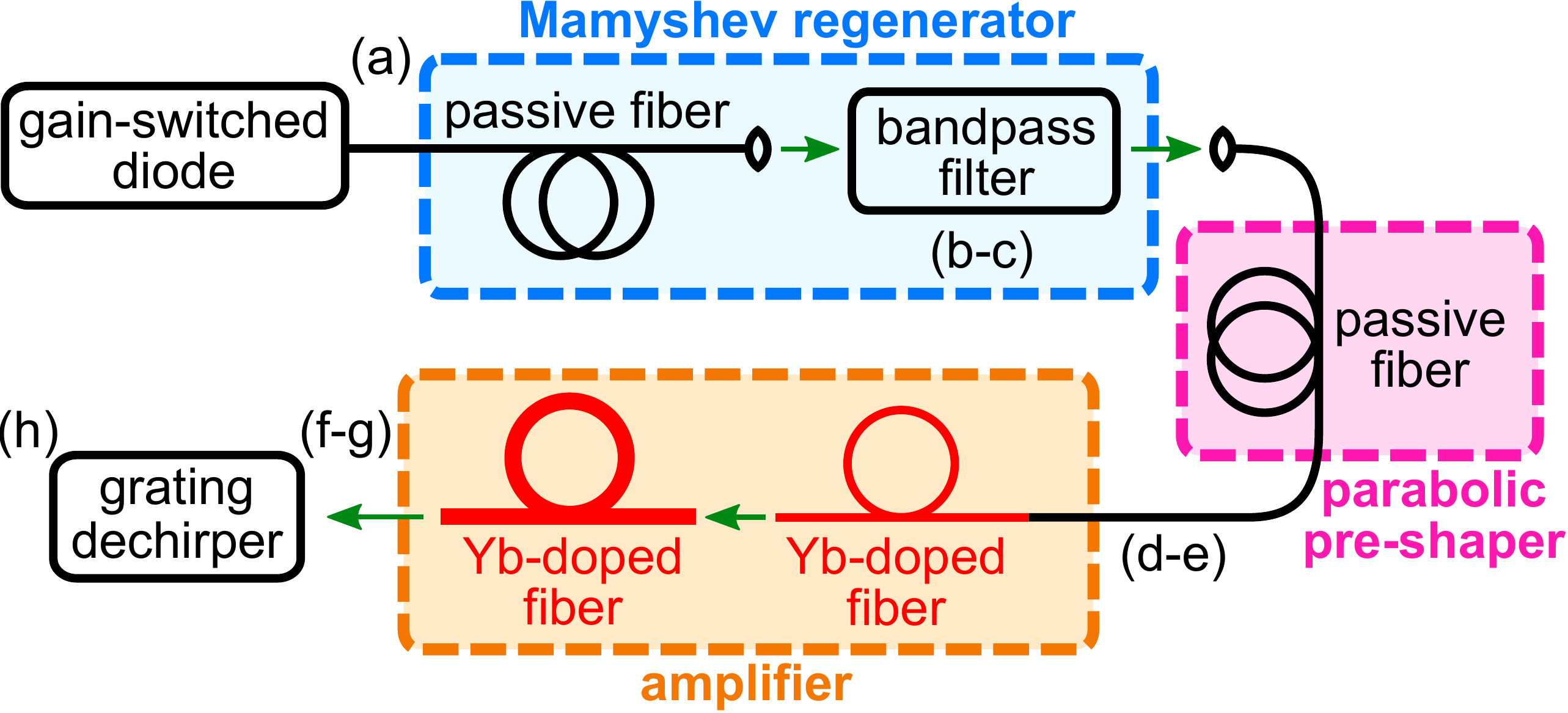}
\caption{Schematic of the experimental system.  Labels correspond to panels in Figure \ref{fig:sim}.}
\label{fig:setup}
\end{figure}

To illustrate the nonlinear processes we exploit, we numerically simulate a representative system.  The results are shown in Figure \ref{fig:sim}, corresponding to the locations labeled in Figure \ref{fig:setup}.  We start with an experimentally-motivated model of a pulse from a GSD, comprising a 2.5-nJ, 10-ps, coherent, Gaussian core superimposed on a 2-nJ, broad, incoherent pedestal (Fig. \ref{fig:sim}a).  Various real or artificial saturable absorbers can differentiate between these two pulse components to some degree.  However, using a Mamyshev regenerator offers a number of particular advantages, including strong and uniform suppression of the incoherent pedestal and a relatively flat (in the ideal case) transmittance for higher-energy pulses.  This stands in contrast to alternatives such as nonlinear optical loop mirrors and nonlinear polarization evolution, which feature oscillatory transfer functions, or semiconductor saturable absorbers, which offer limited modulation.  In addition to its uses as a pulse regenerator \cite{Mamyshev1998}, Mamyshev regeneration has also been demonstrated experimentally as a modelocking mechanism \cite{Regelskis2015,Samartsev2017,Liu2017} and numerically to increase the coherence of a field \cite{Pitois2007}.  The key advance we report here is to use a Mamyshev regenerator as a simultaneous pulse compressor and coherence discriminator, and to subsequently use these properties to enable further nonlinear pulse shaping.

In our simulated Mamyshev regenerator, the pulse first propagates through 63 meters of passive fiber (Nufern PM980-XP).  Self-phase modulation and optical wavebreaking act on the coherent, high-intensity core to generate a series of new spectral components.  Meanwhile, the less-intense pedestal spectrally broadens negligibly.  Filtering a portion of the newly-created bandwidth therefore isolates part of the coherent component, producing a 0.1-nJ, 3-ps pulse within 40\% of its transform-limited duration (Fig. \ref{fig:sim}b-c).  Importantly, the pulse is clean and more coherent, making it a good candidate for amplification by parabolic pre-shaping \cite{Pierrot2013}.  In this technique, an optimized length of normally-dispersive, passive fiber is used to parabolically shape the pulse \cite{Finot2007}, which is then amplified in the highly nonlinear regime.  The parabolic pulse shape converts nonlinear phase into a linear frequency sweep, permitting compression ratios of 30 or more without compromising the pulse quality \cite{Pierrot2013,Fu2017}.  Here, we simulate shaping the pulse in 33 meters of passive fiber (Nufern PM980-XP; Fig. \ref{fig:sim}d-e), before amplifying it first in a core-pumped fiber amplifier (1.4 m Nufern PM-YSF-HI) and then in a 2.3-m double-clad fiber amplifier.  The latter is modeled after chirally coupled core (3C\textsuperscript{\textregistered}) fiber, which uses a large, 34-$\mu$m core diameter to manage high-energy amplification while a helical, secondary core maintains single-mode behavior \cite{McComb2014}.  While this is an attractive feature, our system is equally compatible with conventional fiber amplifiers.  The pulse can be amplified to 3 $\mu$J (Fig. \ref{fig:sim}f-g) before distortions from gain narrowing and stimulated Raman scattering become noticeable, and dechirps to 140 fs in a realistic grating compressor (Fig. \ref{fig:sim}h).  It is worth remarking that if amplification by parabolic pre-shaping is applied directly to the GSD pulse, the pedestal will remain incompressible, causing only a fraction of the total energy to contribute to the peak power.  Furthermore, studies indicate that Raman scattering will prevent a 10-ps pulse from being compressed to significantly less than 300 fs \cite{Fu2017}.  Thus, the combination of Mamyshev regeneration and amplification by parabolic pre-shaping is critical to obtaining such short, clean pulses from a typical GSD.

\begin{figure}[htb]
\centering
\includegraphics[width=0.5\linewidth]{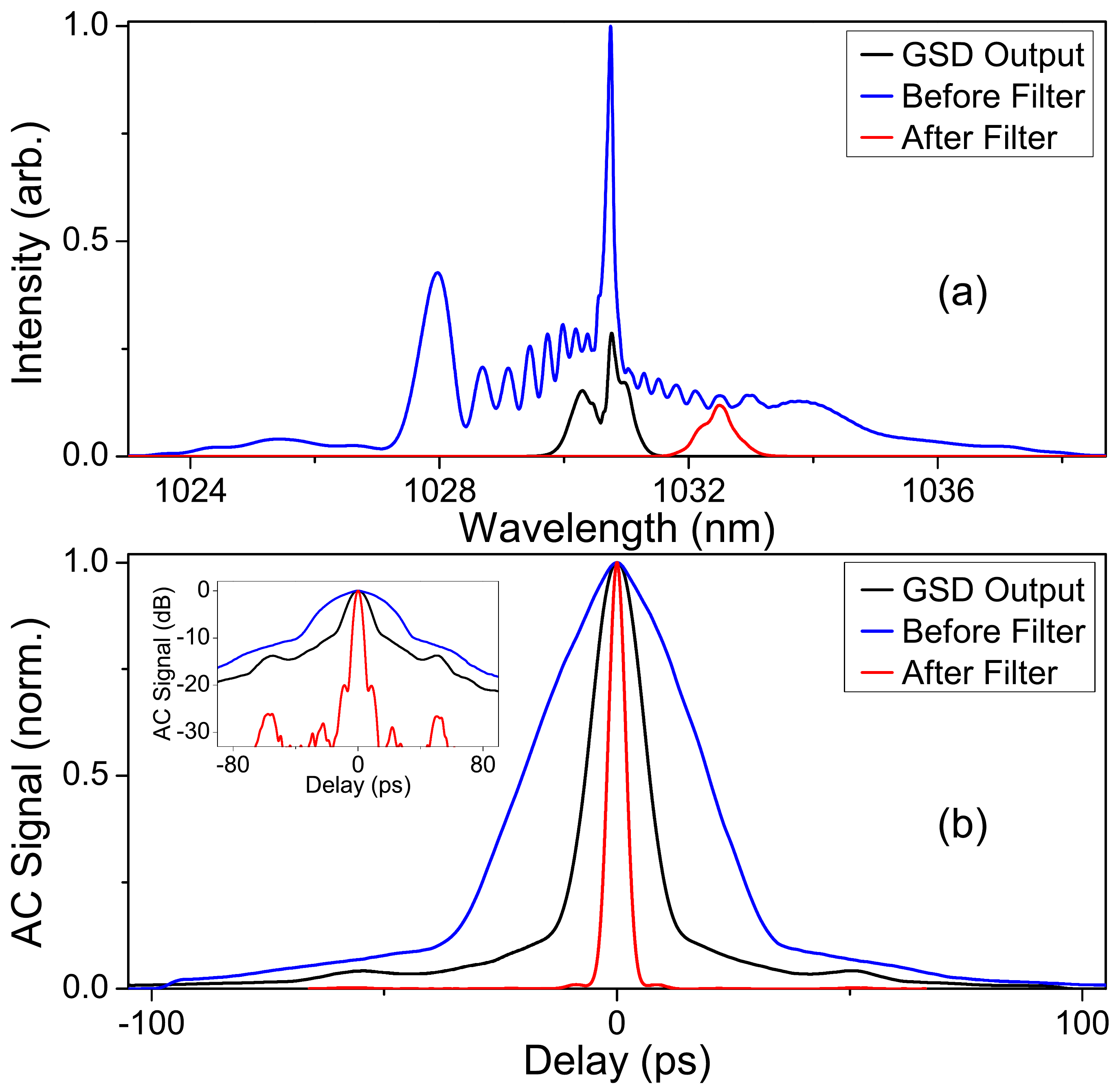}
\caption{(a) Spectra (scaled for visibility) and (b) autocorrelations of the pulses at the GSD output, after spectral broadening, and after filtering.  Inset: logarithmic scale.}
\label{fig:filter}
\end{figure}

Encouraged by these results, we design an equivalent experimental system along the lines of Figure \ref{fig:setup}.  Our starting point is a commercial GSD (OneFive Katana-10 LP), which produces pulses similar to those numerically modeled.  The repetition rate is continuously tunable from 50 kHz to 10 MHz, and is set at 1 MHz for most of what follows.  The autocorrelation at the diode output reveals a prominent, 14-ps peak on top of a broad, complicated background.  As in simulations, we use a Mamyshev regenerator with a tunable bandpass filter to isolate part of the coherent component.  The filter characteristics are experimentally determined, and will be explained later.  Figure \ref{fig:filter} depicts the resulting effects on the pulse.  As expected, the Mamyshev regenerator strips away the incoherent pedestal, and we observe primarily a clean, Gaussian autocorrelation.  The 0.1-nJ filtered pulses are 3 ps long, within 40\% of the transform limit.

\begin{figure}[htb]
\centering
\includegraphics[width=0.5\linewidth]{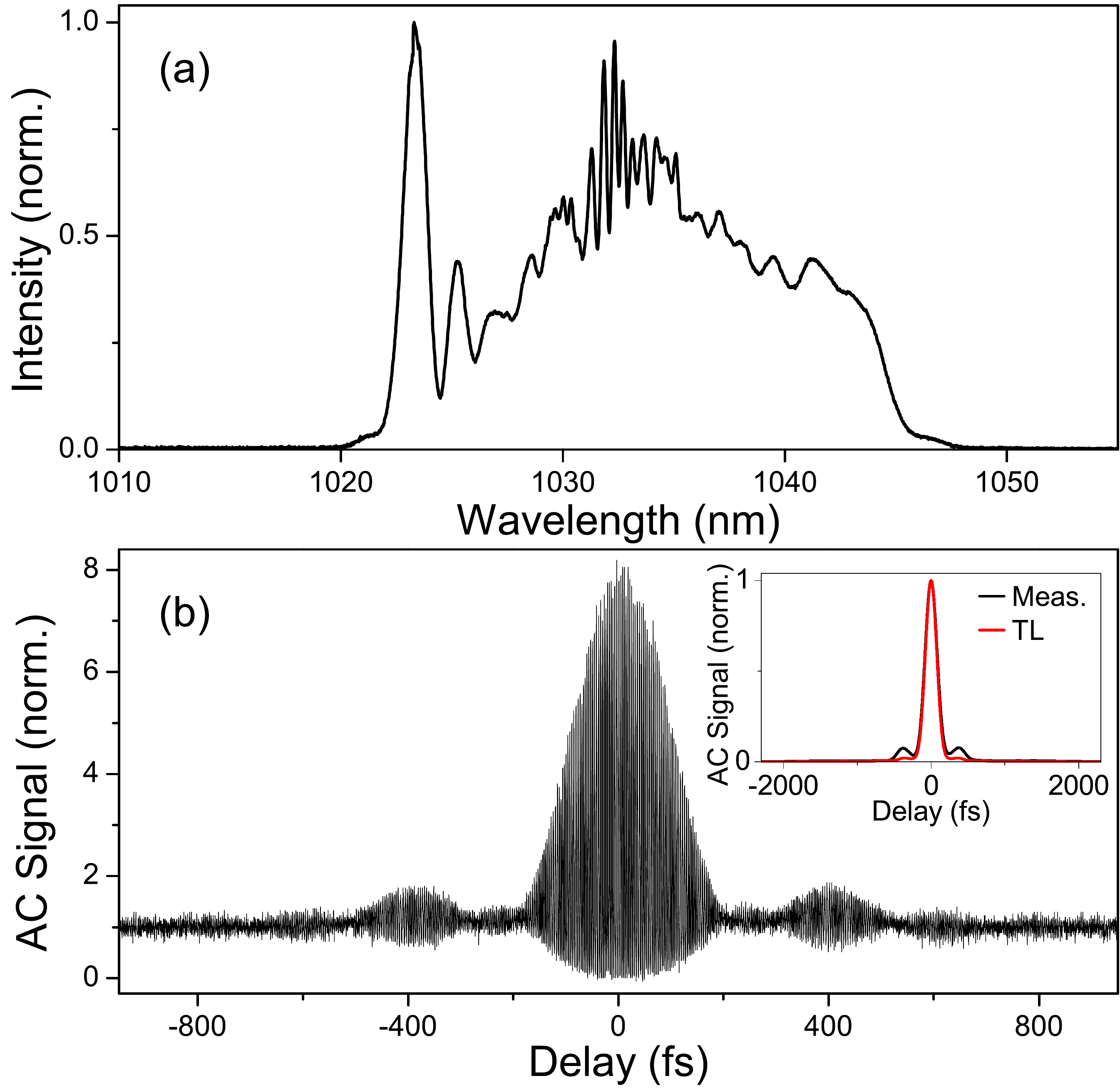}
\caption{(a) Spectrum and (b) interferometric autocorrelation of the output pulses (inset: measured intensity autocorrelation and its transform-limited counterpart).}
\label{fig:output}
\end{figure}

Following amplification by parabolic pre-shaping and dechirping, we measure the pulses shown in Figure \ref{fig:output}.  After dechirping the 3.2-$\mu$J pulses, we obtain a pulse energy of 2.4 $\mu$J.  The modulated and greatly-broadened spectrum reflects the prominence of nonlinearity in the amplifier.  Nevertheless, it is evident that our system efficiently converts nonlinear phase to a cleanly-dechirpable, quadratic phase.  Autocorrelations (Fig. \ref{fig:output}b) show a well-compressed pulse with an inferred duration of 140 fs (using a calculated deconvolution factor of 1.35).  The intensity autocorrelation (inset) confirms this result, and shows that our pulses are very near the 135-fs transform limit (red trace).  Despite the partially coherent nature of the GSD output, the 8:1 ratio of the interferometric autocorrelation indicates a mostly coherent pulse, affirming the role of the Mamyshev regenerator in promoting coherence.  When we launch a small fraction of the dechirped pulses into passive fiber, we observe spectral broadening consistent with simulations where the peak power of the full pulse is taken to be 13 MW.  This eliminates the possibility that the pulses contain a significant pedestal or noise burst component.  As an example of the fine control afforded by the GSD, we also characterize the output pulses after changing the repetition rate slightly.  After adjusting the pumps to maintain the same pulse energy, we obtain comparable performance at 0.8 MHz (see Appendix \ref{sec:reprate}).  In our system, we find that the pulse energy is limited by stimulated Raman scattering.  Hybrid systems employing GSDs, fiber nonlinearities, and solid-state or thin disk nonlinear amplification might yield significantly higher energies \cite{Pouysegur2015,Ueffing2016}.

One disadvantage of our system is that the output shows significant pulse-to-pulse fluctuations.  A GSD generates independent pulses from spontaneous emission, making each pulse slightly different.  Subjecting these pulses to cascaded, nonlinear processes magnifies the differences.  Careful design of the Mamyshev regenerator can produce a locally flat transfer function, reducing this effect \cite{Provost2007}.  It is this consideration which guides our choice of the filter characteristics, leading to the filters shown in Figures \ref{fig:sim}(b-c) and \ref{fig:filter}.  Even so, RF spectral measurements \cite{Gross2002} indicate that the root-mean-square (rms) uncorrelated energy jitter downstream of the bandpass filter is still 2\%.  Our system also displays non-negligible uncorrelated timing jitter (40 ps rms), as is typical of GSD-based systems and is often exacerbated by Mamyshev regeneration \cite{Provost2007}.  Because many ultrafast applications are ultimately sensitive to peak power, we additionally characterize our system's nonlinear noise by measuring the RF spectrum with a two-photon detector.  We find an rms two-photon photocurrent amplitude jitter of 6\%, consistent with the measured 2\% energy jitter.

Although these pulse-to-pulse fluctuations will be too high for some applications, some discussion of the noise is warranted.  Firstly, the timing jitter in a modelocked oscillator differs in character from that in a GSD.  In the former, the pulse arrival time is highly correlated between subsequent roundtrips, and the timing jitter manifests as a gradual drift over some observation period \cite{Jiang2002}; in the latter, the timing jitter derives from spontaneous emission noise preceding pulse formation, and is both uncorrelated from pulse to pulse and independent of the observation period.  As a result, the 40-ps timing jitter we measure should not be compared directly to that of a modelocked oscillator: over a short observation period, the modelocked oscillator will accumulate much less jitter, while over longer timescales, an oscillator lacking active timing stabilization will drift more widely than a GSD (see Appendix \ref{sec:noise} for further discussion).  Secondly, the estimated 13-MW peak power represents a pulse ensemble average, making it a valid performance metric for noise-tolerant applications.  We furthermore measure the single-pulse spectra using the dispersive Fourier transform technique \cite{Tong1997}, and confirm that the distributions of the pulses and their energies are not heavy-tailed, nor are the averages strongly skewed by extreme outliers (see Appendix \ref{sec:noise}).  Finally, techniques such as self-seeding \cite{Lundqvist1983,Schell1994,Vu2008} and external injection \cite{Andersson1982,Seo1995} have been thoroughly investigated as ways of curtailing shot-to-shot fluctuations in GSDs, reducing timing jitter, or even creating interpulse coherence \cite{Teh2014}.  We expect that implementing these approaches in a similar system to ours will yield comparable noise reductions, making such systems viable alternatives to modelocked oscillators even for noise- or coherence-sensitive applications.

In conclusion, we present a system which generates microjoule-scale, 140-fs pulses from a GSD.  We estimate the peak power to be 13 MW, an order of magnitude higher than previous systems based on GSDs.  Attaining this level of performance has historically only been possible by amplifying a modelocked oscillator; by breaking with this trend, our system offers applications an unprecedented level of control over the delivered repetition rate in a highly robust format.  While pulse-to-pulse amplitude fluctuations on the order of a few percent remain a concern, future work using known techniques may resolve this issue, potentially allowing GSD systems in certain fields to rival or even surpass modelocked oscillator technology.

\section*{Funding Information}
National Institutes of Health (NIH) (EB002019); National Science Foundation (NSF) (ECCS-1306035, ECCS-1609129); National Science Foundation Graduate Research Fellowship Program (DGE-1650441); Natural Sciences and Engineering Research Council (NSERC) (CGSD3-438422-2013).

\section*{Acknowledgments}
The authors appreciate insightful feedback from an anonymous reviewer.  They also thank nLIGHT for donating the 3C\textsuperscript{\textregistered} fiber used in these experiments.

\appendix

\section{Repetition rate flexibility} \label{sec:reprate}

\begin{figure}[!ht]
\centering
\includegraphics[width=0.5\linewidth]{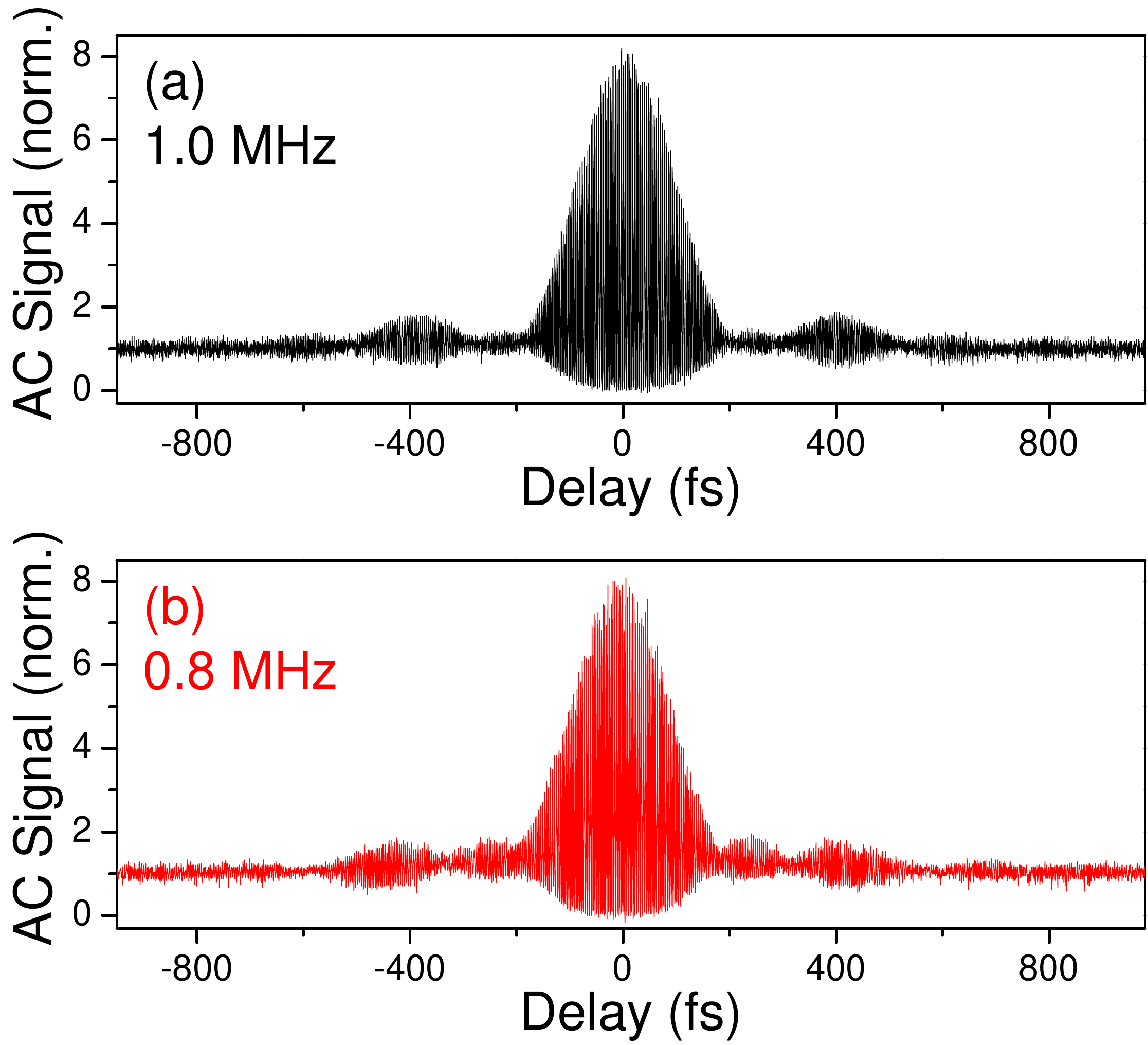}
\caption{Comparison of the autocorrelations obtained at (a) 1.0 MHz and (b) 0.8 MHz.}
\label{fig:reprate}
\end{figure}

The gain-switched diode can be electronically triggered at arbitrary repetition rates.  Provided the amplifiers are adjusted to maintain the same pulse energy in each stage, the nonlinear evolution remains unchanged, and the same level of performance can be maintained.  By way of example, Figure \ref{fig:reprate} depicts the pulses obtained at 0.8 MHz.  Slight differences with the 1 MHz results presented in the main text can be observed, most likely due to the amplifier powers being imperfectly adjusted.  However, the main features remain unchanged: the pulse energy at 0.8 MHz is still 2.4-$\mu$J, and the pulses still dechirp to 150 fs (roughly 10\% longer than the transform limit).

\section{Noise characterization} \label{sec:noise}

\begin{figure}[!ht]
\centering
\includegraphics[width=0.7\linewidth]{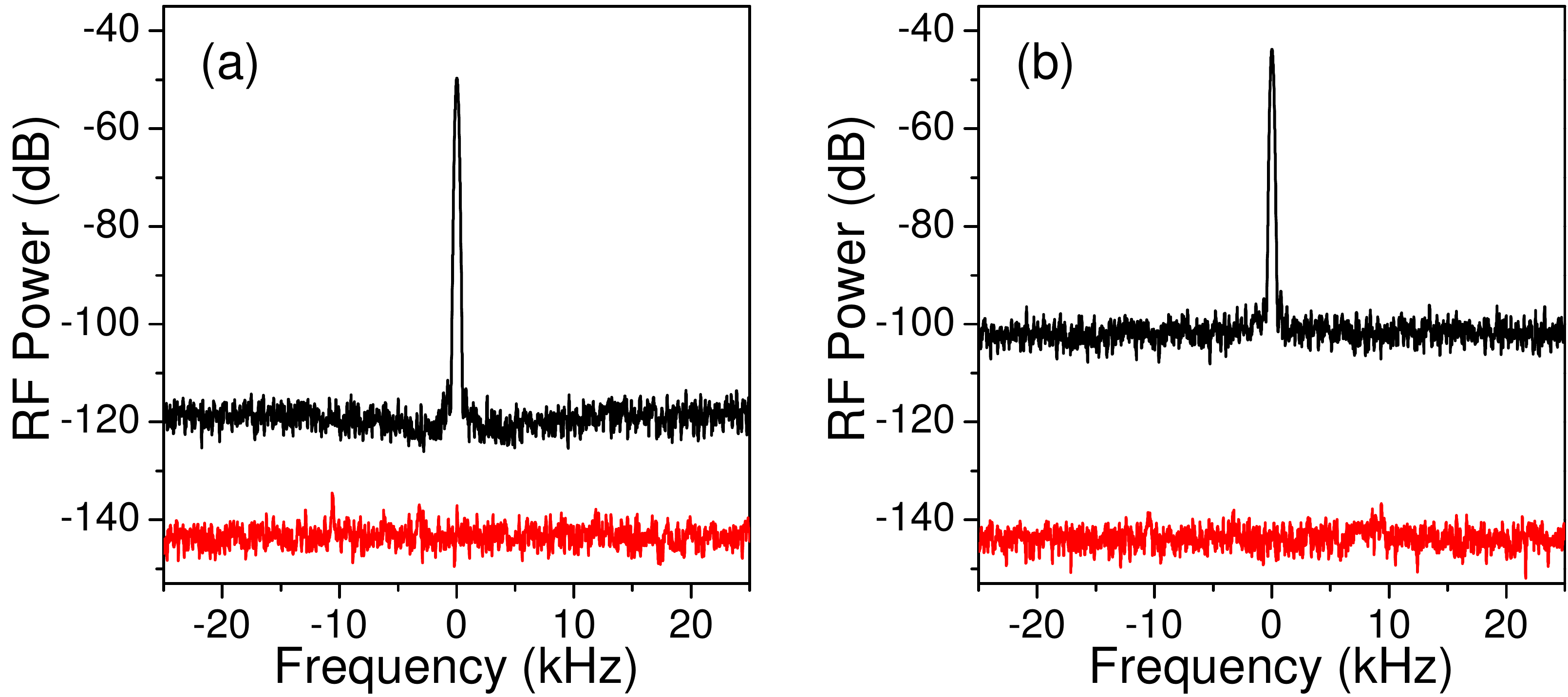}
\caption{Measured RF spectra at the system output using (a) a one-photon detector and (b) and two-photon detector.  Both spectra are centered at 1 MHz, and have a resolution bandwidth of 200 Hz.  Corresponding noise floors are shown in red.}
\label{fig:rf}
\end{figure}

We quantitatively assess the pulse-to-pulse fluctuations using RF spectral intensity measurements.  Figures \ref{fig:rf}(a) and \ref{fig:rf}(b) show the RF spectra observed using a one-photon detector and a two-photon detector, respectively, each measured at the 1-MHz fundamental with a resolution bandwidth of 200 Hz.  In both cases, the spectra consist primarily of the expected comb line and a uniform background.  Following the formalism presented in \cite{Gross2002}, we interpret the latter predominantly as uncorrelated, pulse-to-pulse amplitude noise, which is consistent with theoretical expectations for a gain-switched diode-based source \cite{Schell1994}.  The relative magnitude of this noise can be derived from the ratio between the peak of the comb line and the uniform background, and is 2\% for the one-photon case (corresponding to energy jitter) and 6\% for the two-photon case (corresponding to jitter in the two-photon photocurrent, which is proportional to the product of energy and peak power).

\begin{figure}[!ht]
\centering
\includegraphics[width=0.4\linewidth]{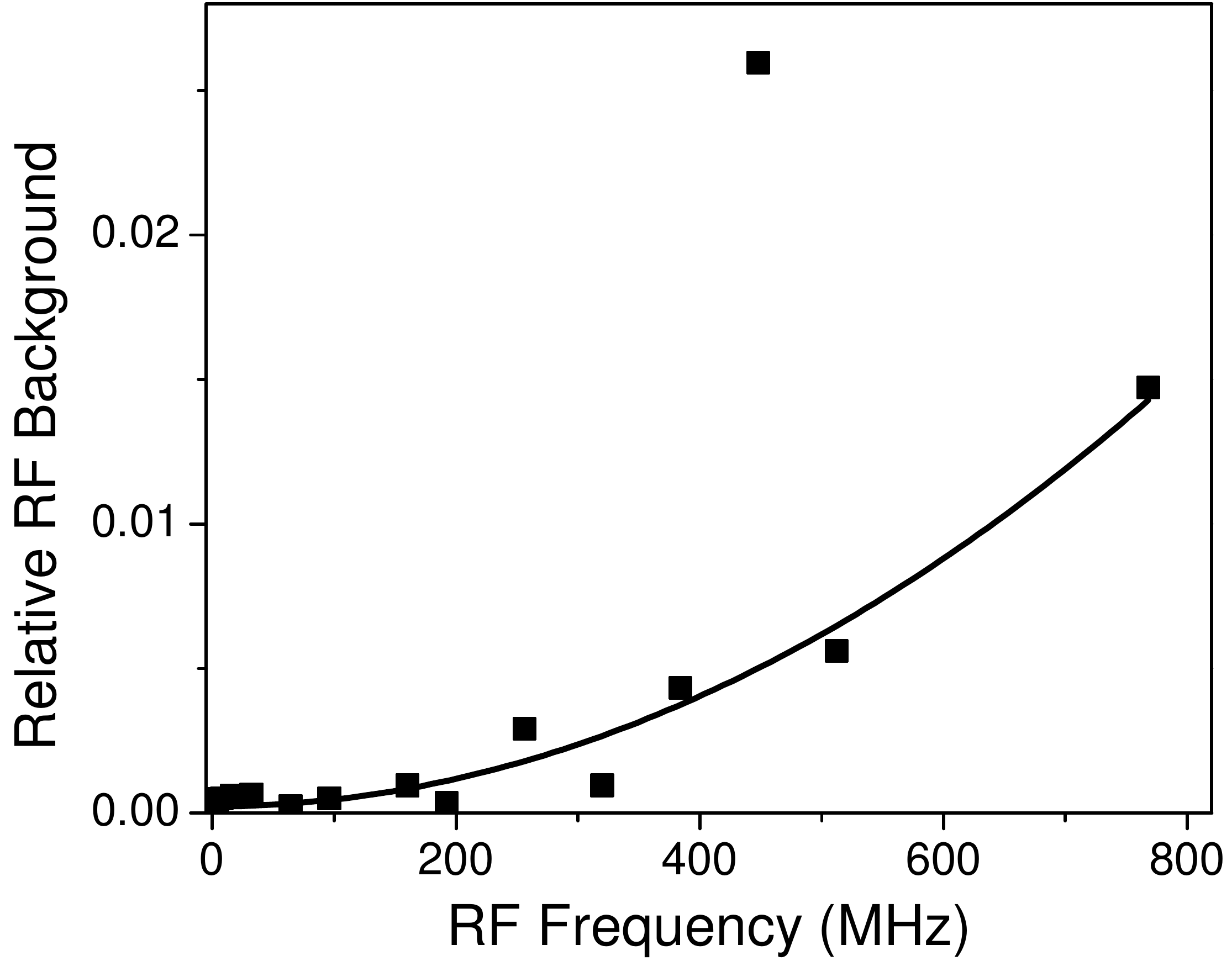}
\caption{Dependence of the RF spectrum background (relative to the comb tooth amplitude) on the RF frequency, measured here at several harmonics of the 1-MHz fundamental.  The solid line represents a quadratic fit as described in the text, with the outlier at 448 MHz omitted.}
\label{fig:timing}
\end{figure}

We determine the timing jitter by measuring the RF spectra centered at harmonics of the repetition rate.  As we look at higher RF frequencies, the overall shape of the RF spectrum remains similar to that shown in Figure \ref{fig:rf}, but the relative background level gradually increases.  This growth obeys a quadratic dependence that is a signature of pulse-to-pulse timing jitter (Fig.~\ref{fig:timing}).  The background near a given harmonic remains uniform, rather than forming a pedestal around the harmonic; we therefore interpret this as uncorrelated timing jitter, as opposed to the correlated variant observed in most modelocked systems.  Keeping with the same formalism as before, we fit the relationship between the relative background and the RF frequency with a quadratic function, omitting a clear outlier at 448 MHz which we attribute to experimental error.  Direct comparison between this fit and equation 3 in \cite{Gross2002} (assuming negligible correlated pulse-to-pulse noise) lets us extract a timing jitter of $40 \pm 20$ ps.  We remark that this method for measuring timing jitter is equivalent to that used by \cite{Schell1994} if amplitude jitter and correlated pulse-to-pulse noise are both negligible.

\begin{figure}[htb]
\centering
\includegraphics[width=0.5\linewidth]{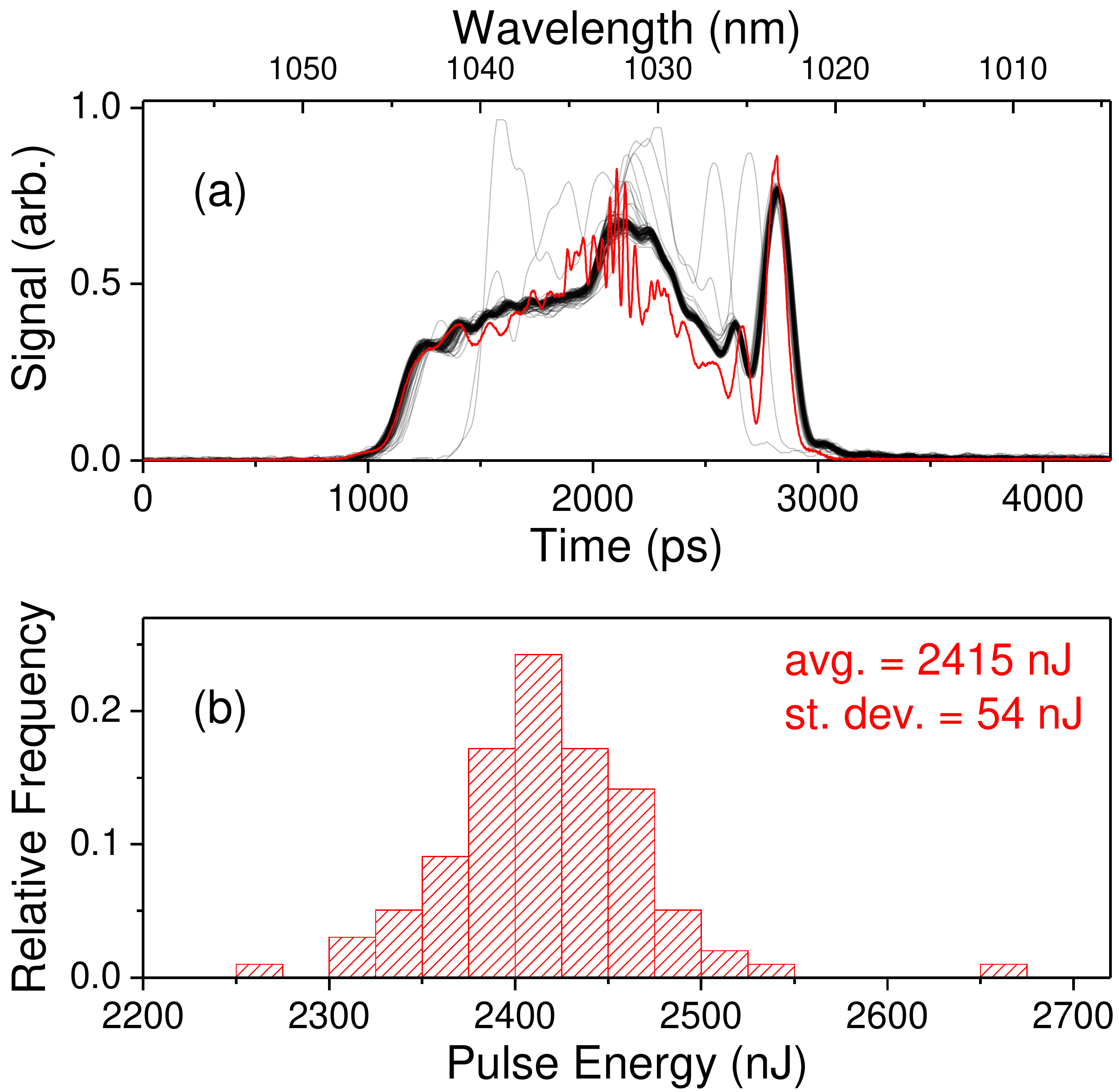}
\caption{(a) Dispersive Fourier transform spectral measurements for 100 consecutive pulses.  Red: average spectrum measured with an optical spectrum analyzer, for comparison.  (b) Histogram of corresponding pulse energies.}
\label{fig:dft}
\end{figure}

As an additional sanity check, both the energy jitter and the timing jitter can be independently estimated from dispersive Fourier transform measurements \cite{Tong1997}.  We disperse the pulses in roughly 5.5 km of standard, step-index fiber (SMF28).  Because this fiber is not strictly single-mode at 1030 nm, a segment of single-mode fiber (Hi1060) is spliced to the end to filter out higher-order modes.  We then directly measure the resulting pulse train with a fast photodetector and a 25-GHz oscilloscope with a long record length.  A single sweep of the oscilloscope is used to measure a series of 100 consecutive pulses while preserving the true timing jitter.  The pulses are overlaid using the nominal repetition rate in Figure \ref{fig:dft}, where it is evident that only two of the 100 pulses differ markedly from the average in their spectral shape.  These two outliers do not noticeably skew the average spectrum (as measured by an optical spectrum analyzer; red line in Fig. \ref{fig:dft}a), nor do they correspond to extreme energies in the single-pulse energy distribution (Fig. \ref{fig:dft}b).  Computing the time-integrated photodetector signal and the center-of-mass arrival time for each pulse, we find standard deviations of 2\% and 30 ps, respectively, in reasonable agreement with the energy jitter and timing jitter derived from RF spectral analysis.

It is important to note that this kind of timing jitter derives from a fundamentally different cause than the kind typically seen in modelocked oscillators.  In the case of the latter, the jitter is correlated from one roundtrip to the next, causing the pulse's arrival time to drift gradually and continuously over time.  Over sufficiently short observation periods, the accumulated timing jitter is negligible.  Over longer observation periods, however, if active stabilization techniques are not used, the timing jitter can reach picosecond scales \cite{Rodwell1989,Crooker1996}.  Timing jitter in modelocked oscillators therefore depends on the observation period (or, equivalently, on the range of RF frequencies over which the noise is integrated).  By contrast, timing jitter in a gain-switched diode is uncorrelated from pulse to pulse.  Each new pulse forms from spontaneous emission noise, and so the arrival times of any two pulses are independent of one another.  Following from the statistics of the initial noise, the arrival times do nonetheless have a statistical distribution.  The measured timing jitter represents the width of this distribution, and is independent of the observation time.  A comparison of timing jitter in modelocked oscillators and gain-switched diodes is therefore meaningless without specifying an observation period (i.e., a frequency range).  Over a short observation time, a gain-switched diode will exhibit greater jitter than an oscillator; over a sufficiently long time, an unstabilized oscillator will undergo an unbounded random walk, while a gain-switched diode will remain anchored within some smaller range of its external trigger \cite{Jiang2002}.  Timing jitter characterization based on a single number obfuscates this difference, and can be misleading.

There certainly exist applications for which a 40-ps uncertainty in the arrival time of a 150-fs pulse will be problematic.  These applications are those that require carrier-envelope offset stabilization (i.e., most frequency comb applications), those that require synchronization of the pulsed source to some other device with sub-40-ps precision, and those that require interactions between subsequent pulses in a pulse train.  In such cases, oscillators using active timing stabilization are typically used to achieve timing jitter on the femtosecond or even attosecond level over typical frequency ranges \cite{Kim2007,Kim2011}.  Nevertheless, the majority of mature applications are insensitive to our observed timing jitter, including many of the most popular applications of high-power, ultrafast fiber lasers. Numerous systems only require synchronization to scanning optics on nanosecond or microsecond time scales.  This is the case in most types of machining and multiphoton microscopy, where there is only one beam present and pulses need only fall within a given dwell time.  Still other applications use a single, master oscillator, and derive multiple pulse trains from a beam splitter (e.g., multi-wavelength sources or optical parametric amplifiers, as well as standard pump-probe experiments).  In these instances, recombination of the same pulse later in the system can render inter-pulse timing jitter a non-issue.

\section{Additional technical details}

Our Mamyshev regenerator filter takes the form of a 1000 l/mm diffraction grating and a fiber-pigtailed collimator.  Together, these effect a tunable, Gaussian bandpass filter with a full-width-half-max of 0.65 nm.  Both the preamplifier and the main amplifier in our system are pumped at 976 nm.  In the results we present, we pump the main amplifier with up to 19 W in the counterpropagating configuration, using free-space coupling for experimental expediency.  Pulse compression is performed using a pair of 1000 l/mm diffraction gratings, which compensate 0.34 ps$^2$ of group delay dispersion.  Simulations suggest that third-order dispersion from the compressor does not significantly degrade the dechirped pulse quality.

Dispersive Fourier transform measurements are performed by dispersing the pulses in roughly 5.5 km of standard, step-index fiber (SMF28).  Because this fiber is not strictly single-mode at 1030 nm, a segment of single-mode fiber (Hi1060) is spliced to the end to filter out higher-order modes.  The resulting pulse train is then directly measured with a fast photodetector and a 25-GHz oscilloscope with a long record length.  A single sweep of the oscilloscope is used to measure a series of 100 consecutive pulses while preserving the true timing jitter.


\end{document}